# A Taylor series solution of the reactor point kinetics equations

David McMahon and Adam Pierson

*Department of Nuclear Safety Analysis,Sandia National Laboratories, Albuquerque, NM 87185-1141*

E-mail address: dmmcmah@sandia.gov

**Abstract**

The method of Taylor series expansion is used to develop a numerical solution to the reactor point kinetics equations. It is shown that taking a first order expansion of the neutron density and precursor concentrations at each time step gives results that are comparable to those obtained using other popular yet more complicated methods. The algorithm developed using a Taylor series expansion is simple, completely transparent, and highly accurate. The procedure is tested using a variety of initial conditions and input data, including step reactivity, ramp reactivity, sinusoidal, pulse, and zigzag reactivity. These results are compared to those obtained using other methods.

## 1. Introduction

The point kinetics equations are a system of coupled ordinary differential equations whose solution gives the neutron density and delayed neutron precursor concentrations in a tightly coupled reactor as a function of time. Typically these equations are solved using a model reactor with at least six delayed precursor groups, resulting in a system consisting of seven coupled differential equations. In general, analytical solutions are not possible due to the complexity of the problem, hence the need for numerical methods. Obtaining accurate results is often problematic because the equations are stiff and with many techniques very small time steps may be required.

In an effort to address these concerns in recent years several new approaches have been proposed. Among the methods that have been discussed are Power Series Solutions (PWS) (Aboanber, 2002, Sathiyasheela 2009), CORE (Quintero-Leyva, 2008), and PCA(Kinard and Allen, 2003). Each of these methods is highly accurate, but they vary widely in complexity of implementation.

In this paper we describe a method that is surprisingly simple while maintaining a desired degree of accuracy. This method, which is fundamentally similar to PWS but is far simpler both in clarity and implementation, is based on taking a Taylor series expansion of the neutron density and delayed precursor functions at each time step. The algorithm is tested using a simple first-order expansion applied to problems involving several conventional reactivity inputs. What's surprising is that a reasonable level of accuracy is achieved even with complex time-dependent reactivity inputs. The solutions obtained are compared with those from other algorithms and with exact solutions where possible.

Several years ago the idea of using a Taylor series to solve the point kinetics equations was dismissed as unworkable, with researchers going so far as to propose Taylor series solutions had no future in the field (Henry, 1971). Later work suggested that a Taylor series could be used, but the author proposed using a complicated transform method to make it work (Mitchell, 1977). However these conclusions may be premature.

We have tested the Taylor series method using several different reactivity inputs and have found the results to be favorable. Conditions tested include step reactivity (prompt sub-critical, critical, and super-critical), ramp, sinusoidal, zigzag, pulse, and reactivity as a function of neutron density. In all cases only a first order expansion was used, resulting in a code that is incredibly simple. Moreover, the method seems to work well for neutron generation times on the order of $\Lambda \sim 10^{-4}$ and $\Lambda \sim 10^{-5}$, indicating it is suitable for use in modeling thermal reactors. In that case, very small time steps are not required.

In the final section, we demonstrate that if the step size is adjusted appropriately, the method works well for neutron generation times on the order of $\Lambda \sim 10^{-7}$, making the algorithm suitable even for modeling fast reactors.

## 2. A Quick Review of the Taylor series algorithm

The time-dependent behavior of small, tightly coupled reactors is well-described by the point kinetics equations (Duderstadt and Hamilton, 1976). Without source and assuming six groups of delayed neutron precursors, these equations take the form

$$\frac{dN(t)}{dt} = \frac{\rho(t) - \beta}{\Lambda} N(t) + \sum_{i=1}^{6} \lambda_i C_i(t) \tag{1}$$

$$\frac{dC_i(t)}{dt} = \frac{\beta_i}{\Lambda} N(t) - \lambda_i C_i(t) \tag{2}$$

The Taylor series expansion of the neutron density can be written as

$$N(t+h) = N(t) + h\frac{dN}{dt} + \frac{1}{2!}h^2\frac{d^2N}{dt^2} + \cdots \tag{3}$$

Similarly, for the delayed precursor concentration we have

$$C_i(t+h) = C_i(t) + h\frac{dC_i}{dt} + \frac{1}{2!}h^2\frac{d^2C_i}{dt^2} + \cdots \tag{4}$$

Taking only terms to first-order, using Eq.(1) we can write an expression that can be used to find the neutron density at a later time $N(t + h)$ from the neutron density at the earlier time $N(t)$

$$N(t + h) = N(t) + h\,\frac{\rho(t) - \beta}{\Lambda}\,N(t) + h\,\sum_{i=1}^{6}\lambda_i C_i\,(t) \qquad (5)$$

Each delayed neutron precursor can be calculated using Eq.(2) and Eq.(4), again only taking terms to first-order we find

$$C_i(t + h) = C_i(t) + h\,\frac{\beta_i}{\Lambda}N(t) - h\,\lambda_i C_i(t) \qquad (6)$$

Hence at each time step the algorithm proceeds as follows. Compute the right-hand side of the point kinetics equations (Eq. (1) and Eq.(2)) using the neutron density and precursor concentrations from the previous time step. Then multiply the result by the time-step size $h$. This surprisingly simple approach gives the neutron density and delayed precursor concentration at each time step with accuracies that rival those obtained with far more complex algorithms. We now consider computational results obtained from this method.

### 3. Numerical Results

In the following we consider the several reactivity inputs: step, ramp, sinusoidal, pulse, zigzag, and as a function of neutron density. To test the algorithm a code was written using the interpreted language *Python*. The code is extremely simple and could easily be translated into FORTRAN, C/C++ or MATLAB.

We begin by considering a step reactivity insertion with $\beta = 0.007$. First two cases were considered, a prompt subcritical step reactivity with $\rho = 0.003$, and a prompt critical step reactivity with $\rho = 0.007$. The step size taken was $h = 0.001$. For comparison, we chose the CORE algorithm (Quintero-Leyva, 2008) and PCA (Kinard and Allen, 2004). The data are presented in Tables 1 and 2 along with exact results, obtained from Chao and Attard (1985). All seem to be in agreement.

The following input parameters were used

$\lambda_i$ = 0.0127, 0.0317, 0.155, 0.311, 1.4, 3.87

$\beta_i$ = 0.000266, 0.001491, 0.001316, 0.002849, 0.000896, 0.000182

$\Lambda = 0.00002$

**Table 1**

Results obtained for prompt subcritical step reactivity $\rho = 0.003$

| Time (s) | CORE | PCA | Taylor | Exact |
|---|---|---|---|---|
| $t = 1$ | 2.2098 | 2.2098 | 2.2099 | 2.2098 |
| $t = 10$ | 8.0192 | 8.0192 | 8.0192 | 8.0192 |
| $t = 20$ | 28.297 | 28.297 | 28.297 | 28.297 |

**Table 2**

Results obtained for prompt critical step reactivity $\rho = 0.007$

| Time (s) | CORE | PCA | Taylor | Exact |
|---|---|---|---|---|
| $t = 0.01$ | 4.5088 | 4.5088 | 4.5086 | 4.5088 |
| $t = 0.5$ | $5.3458 \; X \; 10^3$ | $5.3459 \; X \; 10^3$ | $5.3447 \; X \; 10^3$ | $5.3459 \; X \; 10^3$ |
| $t = 2$ | $2.0600 \; X \; 10^{11}$ | $2.0591 \; X \; 10^{11}$ | $2.0566 \; X \; 10^{11}$ | $2.0591 \; X \; 10^{11}$ |

In the supercritical case, the results are a bit murky. Mathematica, CORE, and the Taylor series method give radically different results than those reported by Kinard and Allen. To clear the waters, a hand solution was derived for one precursor group with the following data:

$\lambda = 0.077$, $\beta = 0.007$, $\Lambda = 0.00002$, $\rho = 0.008$

The solution obtained analytically is found to be

$$N(t) = 7.8338 \, e^{50.5325 \, t} - 6.8333 \, e^{-0.6095 \, t}$$

Numerical solutions calculated using CORE, the Taylor series method, and Mathematica's built-in differential equation solver. These were then compared to values obtained with the analytical solution. The results are shown in Table 3.

**Table 3**

Results obtained for prompt super-critical step reactivity $\rho = 0.008$, single precursor group

| Time (s) | CORE | Mathematica | Taylor | Exact |
|---|---|---|---|---|
| $t = 0.01$ | 6.1296 | 6.1921 | 6.2415 | 6.1929 |
| $t = 0.05$ | 89.5152 | 91.3726 | 91.2466 | 91.3821 |
| $t = 0.1$ | 1244.72 | 1219.69 | 1210.3317 | 1219.80 |
| $t = 0.5$ | $7.3942 \times 10^{11}$ | $7.3610 \times 10^{11}$ | $6.9422 \times 10^{11}$ | $7.3615 \times 10^{11}$ |
| $t = 1.0$ | $6.9504 \times 10^{22}$ | $6.9172 \times 10^{22}$ | $6.1215 \times 10^{22}$ | $6.9177 \times 10^{22}$ |
| $t = 2.0$ | $6.1412 \times 10^{44}$ | $6.1083 \times 10^{44}$ | $4.7598 \times 10^{44}$ | $6.1086 \times 10^{44}$ |

Mathematica agrees pretty well with the exact solution, and while CORE and Taylor don't agree as well as we would like, but they are pretty close. Now let's return to the six-precursor group case with $\rho = 0.008$ using the same parameters as we did with $\rho = 0.003$ and $\rho = 0.007$. In this case, the solutions obtained with CORE, Mathematica, and the Taylor series method agree with each other, but differ substantially with results obtained with PCA and the claimed "exact" solution given in Kinard and Allen. This data is shown in Table 4.

**Table 4**

Results obtained for prompt critical step reactivity $\rho = 0.008$

| Time (s) | CORE | PCA | Mathematica | Taylor | Exact? |
|----------|------|-----|-------------|--------|--------|
| $t = 0.01$ | 6.2029 | 6.0229 | 6.2029 | 6.2080 | 6.0229 |
| $t = 0.5$ | $2.1071 \; X \; 10^{12}$ | $1.4104 \; X \; 10^{3}$ | $2.1071 \; X \; 10^{12}$ | $2.1398 \; X \; 10^{12}$ | $1.4104 \; X \; 10^{3}$ |
| $t = 2$ | $5.2735 \; X \; 10^{46}$ | $6.1634 \; X \; 10^{23}$ | $5.2735 \; X \; 10^{46}$ | $5.6255 \; X \; 10^{46}$ | $6.1634 \; X \; 10^{23}$ |

Now we turn to a ramp reactivity of 0.01\$/sec. In this case the kinetics parameters used in the calculation were

$\lambda_i$ = 0.0127, 0.0317, 0.115, 0.311, 1.4, 3.87

$\beta_i$ = 0.000266, 0.001491, 0.001316, 0.002849, 0.000896, 0.000182

$\Lambda$ = 0.00002

For the Taylor algorithm and CORE, a step size of $h = 0.0001$ was used, while PCA data is that presented in Kinard and Allen in 2004 for $h = 0.01$. Exact data quoted are from Van den Eynde in 2006. These results are shown in Table 5.

**Table 5**

Results obtained for ramp reactivity

| Time (s) | CORE | PCA | Taylor | Exact |
|----------|------|-----|--------|-------|
| $t = 2$ | 1.3382 | 1.3382 | 1.3382 | 1.3382 |
| $t = 4$ | 2.2285 | 2.2278 | 2.2285 | 2.2284 |
| $t = 6$ | 5.5822 | 5.5802 | 5.5823 | 5.5821 |
| $t = 8$ | 42.790 | 42.772 | 42.789 | 42.786 |
| $t = 9$ | 487.61 | 487.35 | 487.52 | 487.52 |

Next we consider the case of sinusoidal reactivity. In this case the kinetics parameters used were

$\lambda_i$ = 0.0124, 0.0305, 0.111, 0.301, 1.14, 3.01

$\beta_i$ = 0.000215, 0.001424, 0.001274, 0.002568, 0.000748, 0.000273

$\Lambda$ = 0.0005

The reactivity was a time-dependent function of the form

$$\rho(t) = \beta \sin\left(\frac{\pi \, t}{5}\right)$$

In this case exact data were not available, so the problem was solved with Mathematica's built in numerical differential equation solver for comparison. For the Taylor algorithm and CORE, a step size of $h = 0.0001$ was used. Fig.1 shows a plot of the neutron density obtained using the Taylor method.

**Table 6**

Results obtained for sinusoidal reactivity

| Time (s) | CORE | Taylor | Mathematica |
|----------|---------|---------|-------------|
| $t = 2$ | 10.1475 | 11.3820 | 11.3738 |
| $t = 4$ | 96.7084 | 92.2761 | 92.5595 |
| $t = 6$ | 16.9149 | 16.0317 | 16.0748 |
| $t = 8$ | 8.8964 | 8.6362 | 8.6551 |
| $t = 10$ | 13.1985 | 13.1987 | 13.2202 |

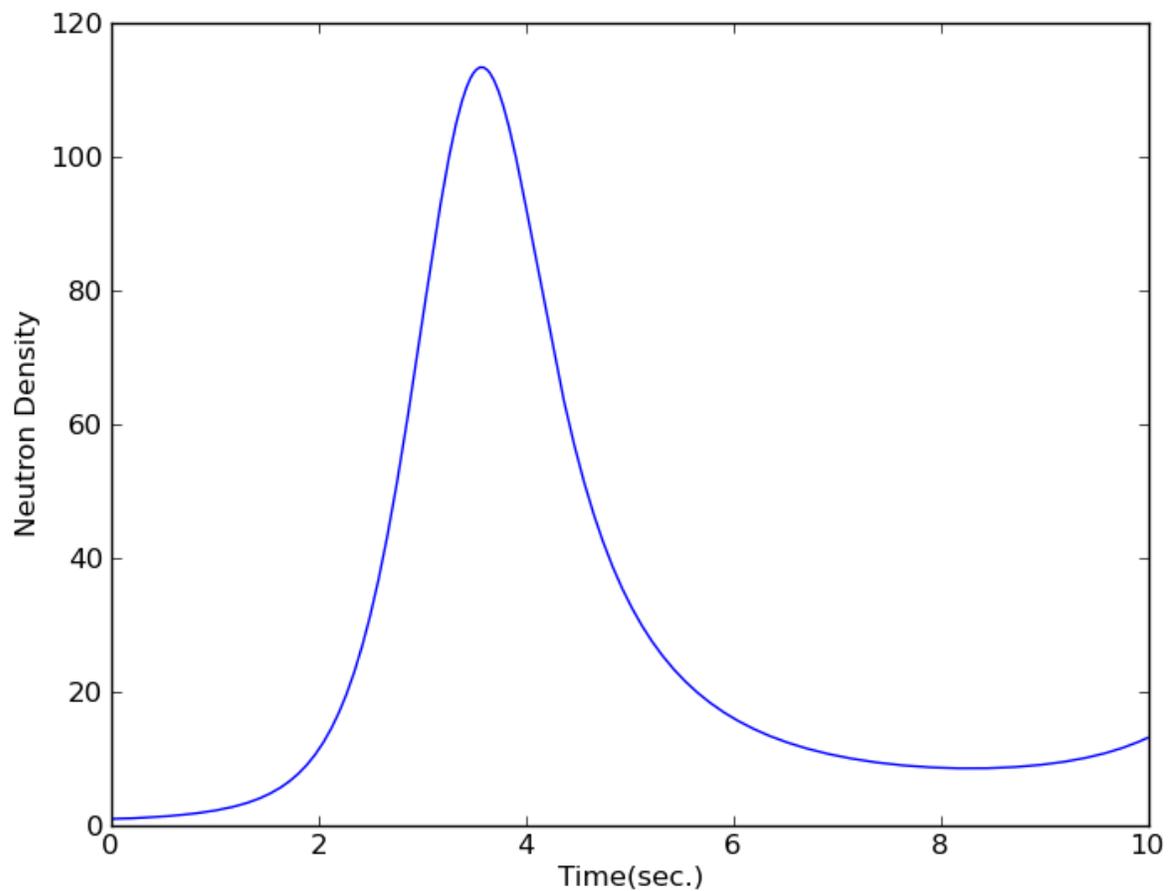

**Fig. 1.** Neutron Density for sinusoidal reactivity calculated with the Taylor series method

The next set considered is a zigzag ramp reactivity. This is defined as follows. A ramp reactivity of $1/s is applied up to 0.5 s. This is followed by a ramp of -$1/s up to 1 s. Up to 1.5 s the reactivity is once again a ramp of $1/s, and thereafter the reactivity is a constant $0.5.

In this case the kinetic parameters used were

$\lambda_i$ = 0.0127, 0.0317, 0.115, 0.311, 1.4, 3.87

$\beta_i$ = 0.000285, 0.0015975, 0.001410, 0.0030525, 0.00096, 0.000195

$\Lambda$ = 0.0005, $\beta$ = 0.0075

An exact solution was not available, so calculations of the neutron density done with the Taylor series method and CORE were again compared to Mathematica using implicit Runge-Kutta (Table 7). **Fig. 2** shows the neutron density calculated with the Taylor series method.

**Table 7**

Results obtained for zig-zag reactivity

| Time (s) | CORE | Taylor | Mathematica |
|----------|---------|---------|-------------|
| $t = 2$  | 8.64539 | 8.6661  | 8.7476      |
| $t = 4$  | 10.0909 | 10.1239 | 10.1931     |
| $t = 6$  | 14.1736 | 14.2184 | 14.3128     |
| $t = 8$  | 20.0586 | 20.1205 | 20.2527     |
| $t = 9$  | 23.8618 | 23.9348 | 24.0916     |
| $t = 10$ | 28.3817 | 28.4678 | 28.6541     |

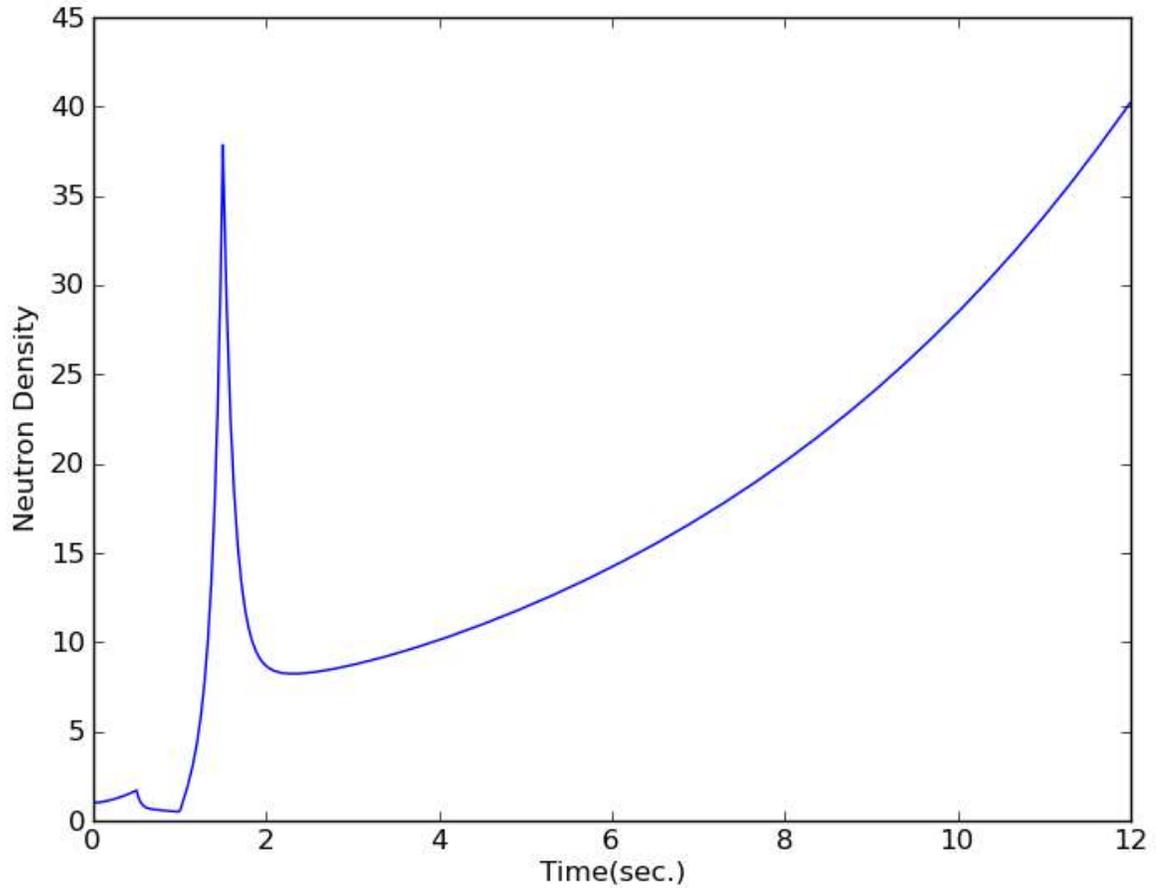

**Fig. 2.** Neutron density for zig-zag reactivity

Now consider a pulse type reactivity of the form

$$\rho = 4\,\beta\,\mathrm{Exp}\,(\,-2t^2\,)$$

for $t < 1$ sec., zero otherwise. The neutron density is illustrated in Fig. 3. This case was testing using a single precursor group, with

$$\lambda = 0.077,\ \beta = 0.006502,$$

For comparison, a solution was found using Mathematica. The results are shown in Table 8. Fig. 3 shows a plot of the neutron density calculated using the Taylor method. This curve has the expected shape for a pulse reactivity.

**Table 8**

Results obtained for pulse reactivity

| Time (s) | Taylor | Mathematica |
|----------|--------|-------------|
| $t = 0.78$ | 15.6791 | 15.6810 |
| $t = 1$ | 9.8670 | 9.8779 |
| $t = 2$ | 1.5629 | 1.6252 |
| $t = 3$ | 1.5629 | 1.6159 |

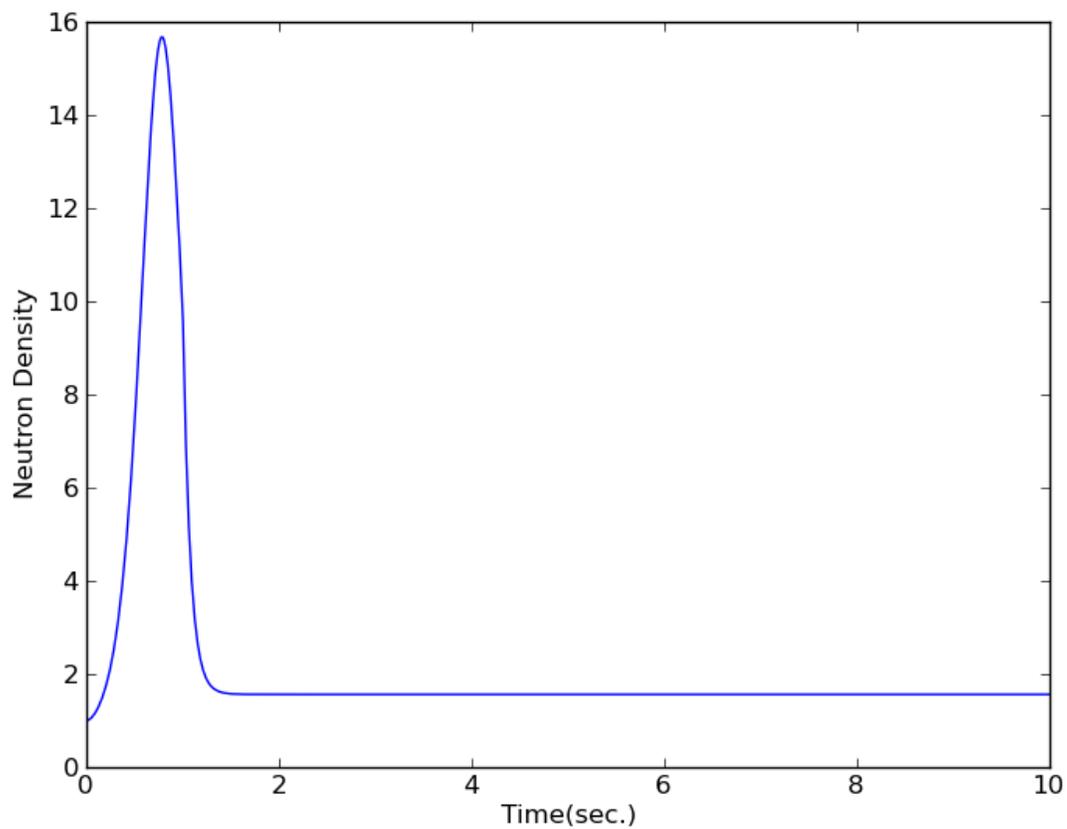

**Fig. 3.** Neutron density calculated using the Taylor method in response to a pulse reactivity.

The final set considered was a reactivity which is a function of neutron density

$\rho = \beta/10 \times N(t)$

The kinetics parameters used are the same as the sinusoidal case. Calculations were done using CORE, the Taylor method, and using Mathematica (implicit Runge-Kutta). The results obtained with the Taylor method and Mathematica are in excellent agreement, as shown in Table 9. A plot of the neutron density is shown in Fig. 4.

**Table 9**

Results obtained for reactivity as a function of neutron density

| Time (s) | CORE | Taylor | Mathematica |
|----------|--------|--------|-------------|
| $t = 2$ | 1.2099 | 1.2091 | 1.2091 |
| $t = 4$ | 1.2886 | 1.2850 | 1.2849 |
| $t = 6$ | 1.3701 | 1.3623 | 1.3623 |
| $t = 8$ | 1.4590 | 1.4452 | 1.4452 |
| $t = 9$ | 1.5073 | 1.4897 | 1.4897 |

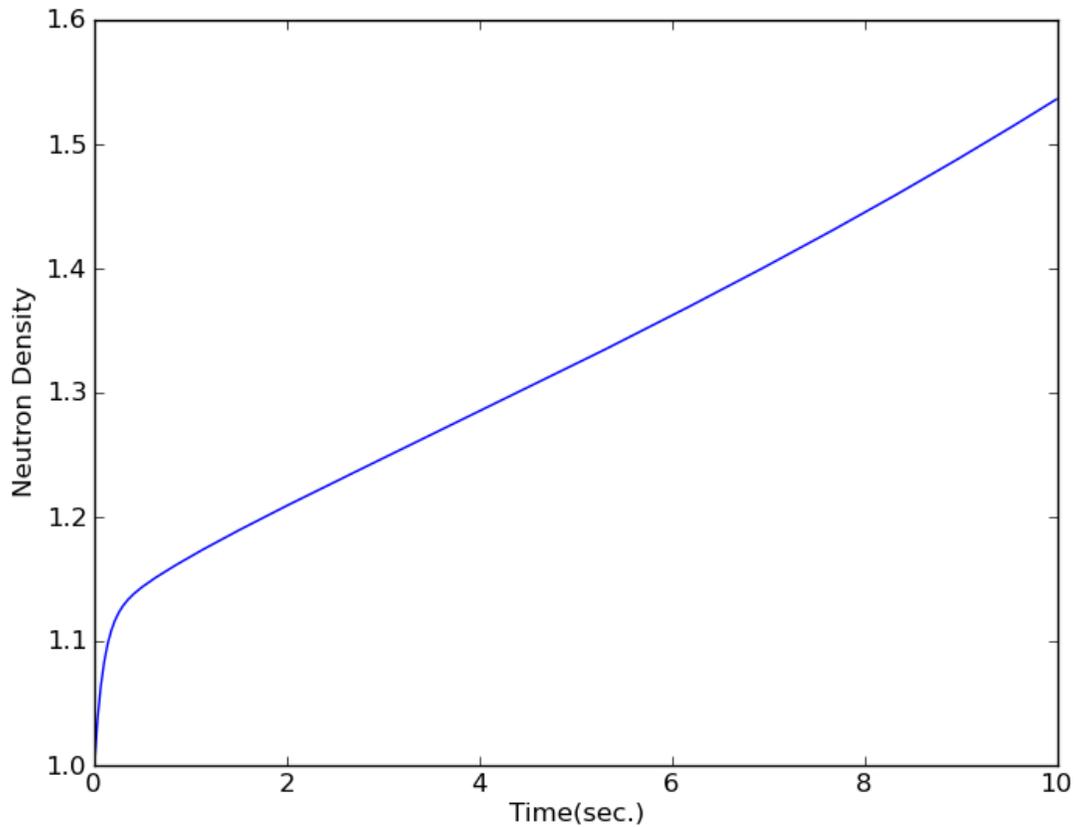

**Fig. 4.** Calculated Neutron density for the case where reactivity is a function of neutron density.

**4. Using the method with small neutron generation times**

A concern with the Taylor series method would be cases where $\Lambda$ is very small, such as with a fast reactor. However results indicate the method is accurate in these cases as well. Following the example in Aboanber (2002) we consider a sinusoidal reactivity

$$\rho(t) = \rho_0 \sin\left(\frac{\pi\,t}{50}\right)$$

where $\rho_0 = 0.005333$. The kinetics parameters used with a single neutron precursor group are $\lambda = 0.077$, $\beta = 0.0079$ with $\Lambda = 10^{-7}$. In order to maintain accuracy, a small step size of $h = 10^{-5}$ was used. This is the downside of the method as currently used, for small $\Lambda$ the step size must be comparatively small requiring longer run times. Nonetheless the results obtained are accurate. For comparison a simulation was run using Mathematica's built in differential equation solver. The results are shown in Table 10, and a plot of the neutron density obtained using the Taylor method is shown in Fig. 5.

**Table 10**

Results obtained for sinusoidal reactivity with with $\Lambda = 10^{-7}$

| Time (s) | Taylor | Mathematica |
|---|---|---|
| $t = 0.01390$ | 1.00059 | 1.00059 |
| $t = 0.10560$ | 1.00452 | 1.00452 |
| $t = 1.00340$ | 1.04619 | 1.04619 |
| $t = 5.15790$ | 1.33979 | 1.33979 |
| $t = 7.96970$ | 1.68691 | 1.68692 |
| $t = 16.00000$ | 4.54887 | 4.54891 |
| $t = 26.33150$ | 25.64396 | 25.64452 |
| $t = 37.22800$ | 60.64332 | 60.64481 |
| $t = 39.24590$ | 61.49032 | 61.49182 |
| $t = 58.41560$ | 31.34570 | 31.34643 |
| $t = 80.68310$ | 13.72866 | 13.72898 |
| $t = 100.00000$ | 15.44012 | 15.45681 |

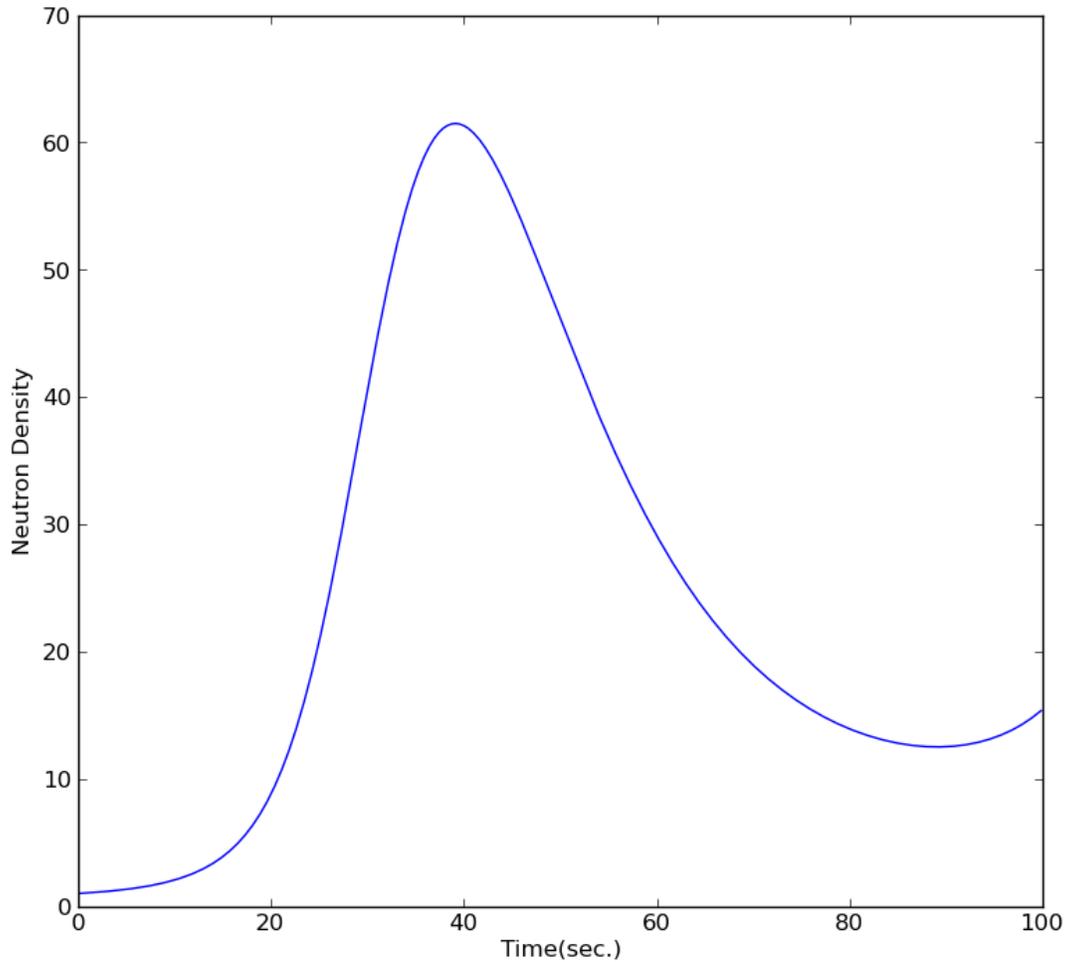

**Fig. 5.** Neutron density obtained for the $\Lambda = 10^{-7}$ case.

### 5. Conclusion and suggestions for future research

The Taylor series method is extremely simple to understand and code, and it generates results that are consistent with other recently published methods. Estimates of CPU time required for the Taylor series method were not carried out because the calculations were done in *Python*, which is an interpreted language. It is expected that the Taylor series method would compare favorably when modeling thermal reactors because fewer calculations per time step are required as compared to other techniques. This may not be true when small neutron generation times are considered, because of the small time steps required. However in all cases the Taylor series method appears to be accurate, even only considering a first order expansion.

Due to the inherent simplicity of the algorithm, it would be worthwhile to develop a compiled code in FORTRAN or C++ to compare the CPU time required to that of other methods. In addition, further studies could be carried out to determine if improvements in accuracy could be obtained by taking more terms in the series and/or smaller time steps or to modify the algorithm to handle the small parameter more effectively (neutron generation time).

**Acknowledgements**

The authors would like to thank Gert Van den Eynde and Barbaro Quintero-Leyva for many helpful discussions.